\documentclass[trackchanges]{aastex701}
\usepackage{mathrsfs}
\usepackage{amsmath}
\usepackage{caption}
\usepackage{xcolor}

\begin{document}

\title{Different Jet Dissipation Mechanisms Underlying the Variability in Blazars}

\author[orcid=0009-0004-7502-7037]{Hong-bin, Tan}
\affiliation{School of Astronomy and Space Science, Nanjing University, 163 Xianlin Avenue,\\
Nanjing 210023, People's Republic of China}
\affiliation{Key Laboratory of Modern Astronomy and Astrophysics,\\
Nanjing University, Ministry of Education, Nanjing 210023, People's Republic of China}
\email{hbtan@smail.nju.edu.cn}  

\author[orcid=0000-0003-1576-0961,gname=Bosque, sname='Sur America']{Ruo-Yu, Liu} 
\affiliation{School of Astronomy and Space Science, Nanjing University, 163 Xianlin Avenue,\\
Nanjing 210023, People's Republic of China}
\affiliation{Key Laboratory of Modern Astronomy and Astrophysics,\\
Nanjing University, Ministry of Education, Nanjing 210023, People's Republic of China}
\affiliation{Tianfu Cosmic Ray Research Center, Chengdu 610000, Sichuan, People's Republic of China}
\email{ryliu@nju.edu.cn}

\correspondingauthor{Ruo-Yu Liu}
\email{ryliu@nju.edu.cn}

\begin{abstract}

Blazars are among the most extreme classes of active galactic nuclei. They are powered by relativistic jets, but the way in which the jet energy is dissipated is still unclear. The flat radio spectrum and the core-shift effect trace the distributions of magnetic fields and relativistic particles along the jet, while variability carries information about time-dependent dissipation. However, a unified framework connecting these observables to the underlying jet physics has been lacking. Here we present a multi-frequency analysis of the prototypical blazar Mrk~501. We model its core-shift measurements, spectral energy distributions (SEDs), and power spectral densities (PSDs) with a conical jet model that conserves magnetic power. The core-shift data localize the radio emitting regions and constrain the electron-density and dissipation-rate profiles along the jet. With a single radial distribution of jet parameters, the model reproduces the core-shift relation and SED, but it underpredicts the observed variability at high radio frequencies and in the optical to $\gamma$-ray bands. We therefore introduce different blob distributions for the inner ($\lesssim$~0.1\,pc) and outer ($\gtrsim$~0.1\,pc) jet regions. With this extended model, the simulated PSDs are consistent with the multiwavelength observations of Mrk~501 during its 2017--2019 low state. This result points to different dissipation behavior in the inner and outer jet. Our study demonstrates that spectro--timing--astrometric jet modeling, which combines SEDs, multiwavelength PSDs, and radio core-shift measurements, can constrain jet stratification and scale-dependent dissipation in blazars.

\end{abstract}

\keywords{\uat{High Energy astrophysics}{739}}


\section{Introduction} 

Blazars are a subclass of active galactic nuclei (AGN) whose relativistic jets are viewed close to the line of sight. They provide useful laboratories for studying extreme physical processes \citep{1995PASP..107..803U}. Their broadband nonthermal emission, from radio to $\gamma$-rays, is strongly amplified by Doppler boosting. This emission is mainly produced by synchrotron and inverse Compton (IC) radiation from accelerated electrons \citep{2008MNRAS.385..283C}, although hadronic processes may also contribute \citep{2013PhRvD..87j3015S}. The high-energy emission is usually associated with compact regions near the central supermassive black hole (SMBH) \citep{1996ApJ...463..555I}. In contrast, the flat radio spectrum (spectral index $\alpha_r\sim 0$) and the frequency-dependent positions of radio cores (the ``core-shift'' effect) indicate an extended jet structure \citep[e.g.][]{2001A&A...367..809K, 2003A&A...410..101K}. These facts suggest that the dissipation process may change from one jet scale to another.

The flat radio spectrum, $F_\nu \propto \nu^{\alpha_r}$ with $\alpha_r\approx 0$, is one of the main features of blazars \citep{2011A&A...536A..15P}. It is commonly interpreted as the superposition of many synchrotron-self-absorbed components produced at different distances along the jet. In the conical jet model of \citet{1979ApJ...232...34B} and later developments \citep{1981ApJ...243..700K, 1985A&A...146..204G, 1980ApJ...235..386M, 2006MNRAS.367.1083K, 2012MNRAS.423..756P}, the magnetic field decreases approximately as $B\propto r^{-1}$. This model links the observed spectrum to the spatial distributions of electrons and magnetic fields. It also predicts that higher-frequency radio emission comes from regions closer to the SMBH, as confirmed by very long baseline interferometry (VLBI) observations \citep[e.g.][]{2011A&A...532A..38S, 2011Natur.477..185H, 2009MNRAS.400...26O}. Thus, radio observations already show that the emitting jet is stratified: the inner regions have stronger magnetic fields and higher particle densities, while the outer regions are more diffuse.

Variability provides complementary information. Blazars vary on timescales from minutes to decades. X-ray and $\gamma$-ray bands often show strong and rapid variability \citep[e.g.][]{2003MNRAS.345.1271V, 2021A&A...655A..93A}, while radio variability is usually weaker and slower \citep[e.g.][]{2021A&A...655A..93A}. The radio variability is also frequency dependent: higher-frequency radio bands tend to show faster and stronger fluctuations \citep[e.g.][]{1985A&A...146..204G, 2023ApJ...944..177K, 2024MNRAS.535.2775V}. This trend is expected if higher-frequency radio photons are emitted closer to the jet base. Power spectral density (PSD) analysis often shows a red-noise shape that flattens at low temporal frequencies \citep[e.g.][]{2019ApJ...885...12R, 2022ApJ...930..157Z}. The physical origin of this behavior is still not clear. While radio spectra and core shifts describe the steady structure of the jet, PSDs probe the time-dependent dissipation process.

A key problem is then how to connect spectral, spatial, and temporal data in one physical picture. To this end, we use the stochastic dissipation model (SDM) developed by \citet{2023MNRAS.526.5054L}. The model divides the jet into many radiating blobs and can simulate multiwavelength light curves and their PSDs. We apply this model to Mrk~501, a nearby ($z=0.034$) high-synchrotron-peaked BL Lac object with extensive multi-epoch data \citep[e.g.][]{2024A&A...685A.117M, 2021A&A...655A..93A}. A multiwavelength campaign in 2009 provided a detailed SED in a typical state \citep{2011ApJ...727..129A}. Another campaign caught the source in a low state with little flaring activity \citep{2023ApJS..266...37A}. These data, together with VLBI core-shift measurements \citep{2010MNRAS.402..259C}, make Mrk~501 a good target for studying scale-dependent dissipation in blazar jets.

The rest of this paper is organized as follows. In Section~2 we describe the model and derive useful analytic relations. In Section~3 we apply the model to Mrk~501 and compare it with the SED, core-shift relation, and multiwavelength PSDs. In Section~4 we discuss the physical implications. Section~5 gives the conclusion.

\section{Model Description}

Following \citet{2023MNRAS.526.5054L}, we describe the low-state emission of a blazar as the sum of radiation from many blobs along the jet. The number of generated blobs per unit time and unit jet length follows a power-law distribution $p(r)\propto r^{-\alpha_{\rm blob}}$, where $r$ is the distance from the SMBH. It is normalized by the total dissipation rate of the jet $\dot{N}=\int p(r)dr$. The blob radius in the comoving frame $R'(r)$ is assumed to be proportional to the transverse radius of the jet $R'(r)=\kappa R_{\rm jet}(r)$, where $\kappa$ is the ratio of the blob radius to the jet radius. For a conical jet, $R'(r)\propto r$. The magnetic field $B'(r)$ and the injected electron luminosity $L^{\prime}_e(r)$ of the blob are also assumed to follow power laws, $B'(r)=B'(r_0)(r/r_0)^{-\alpha_{\rm B}}$ and $L^{\prime}_e(r)=L'_e(r_0)(r/r_0)^{\alpha_{\rm L}}$, where $r_0$ is the distance of the jet base. We set $\alpha_B=1$, which corresponds to approximately conserved magnetic power in a conical jet. The injected electron spectrum is assumed to be $Q'(r,\gamma')=Q'_0(r) {\gamma'}^{-s}$ for $\gamma'_{\rm min} < \gamma' < \gamma'_{\rm max}$, where $\gamma'_{\rm min}$ and $\gamma'_{\rm max}$ are the minimum and maximum electron Lorentz factors. The normalization is $Q'_0(r)=L^{\prime}_e(r)/(m_e c^2 \int{\gamma'}^{1-s}d\gamma')$. To explain the broken power-law spectrum, following \citet{2024MNRAS.529..903T}, we suppose that the maximum electron energy (Lorentz factor) is variable in different blobs in a segment and the blob distribution can be described by $dN/d\gamma'_{\rm max}\propto {\gamma'}_{\rm max}^{-\alpha_{\gamma}}$ for $\gamma'_{\rm max,1} < \gamma'_{\rm max} < \gamma'_{\rm max,2}$, where $\gamma'_{\rm max,1}$ and $\gamma'_{\rm max,2}$ are lower and upper limits of the maximum Lorentz factor. The blobs move inside the jet with bulk Lorentz factor $\Gamma_{\rm bulk}$ in a direction close to the observer's line of sight, leading to Doppler boosting with Doppler factor $\delta_{\rm D}=[\Gamma_{\rm bulk}(1-\beta \cos{\theta_{\rm obs}})]^{-1}$, where $\beta=v_{\rm bulk}/c$ describes the velocity of the blob moving within the jet and $\theta_{\rm obs}$ is the angle between the blob's motion (jet propagation) direction and the observer's line of sight.

Within this model framework, the physical parameters of the emitting regions at different locations along the jet are specified. Based on the electron continuity equation, we numerically compute the evolution of the electron energy distributions and SEDs in all blobs. Assuming that dissipation of the blobs is triggered stochastically at different moments, we can derive the total and distance-resolved light curves, as well as the corresponding time-averaged SEDs over a given time interval.

In this section, we present an analytic derivation of the observational characteristics predicted by the multi-blob model, including the radio spectrum, the core-shift relation, and PSDs. Such an analysis allows for a clearer physical interpretation of the observed features. It also provides useful guidance for the numerical model fitting.

\subsection{Radio Spectrum}

To understand the flat radio spectra, we focus on the radio synchrotron radiation of blazars. In a spherical emission region (blob), synchrotron radiation originating from a power-law electron spectrum is given as \citep[e.g.][]{1979rpa..book.....R, 2009herb.book.....D}
\begin{equation}\label{eq:L_blob_syn}
    L^{\rm syn \ \prime}_{\nu, \rm blob}(r, \nu') = \frac{3^{(s+3)/2}}{2^{(s+1)/2}} C_1(s) \frac{4}{3} c \sigma_T u'_B(r) K'_e(r) \frac{\nu'^{(1-s)/2}}{\nu'_B(r)^{(3-s)/2}},
\end{equation}
where $u'_B(r)$ is the magnetic field energy density at distance $r$, $\nu'_B(r)=e B'(r)/(2\pi m_e c)$ and $C_1(s)$ is a constant related to the spectral index $s$. $K'_e(r)=Q'_0(r)t'_{\rm inj}(r)$, where $t'_{\rm inj}(r)$ is the injection timescale of electrons and is equal to $R'(r)/c$ in this work. In the radio band, the effect of synchrotron self-absorption is unavoidable. Synchrotron emission is suppressed by a factor $\frac{1-e^{-\tau_{\rm SSA}}}{\tau_{\rm SSA}}$, where $\tau_{\rm SSA}$ is the opacity due to synchrotron self-absorption and is given as \citep{2009herb.book.....D}
\begin{equation}\label{eq:tau_ssa}
    \tau_{\rm SSA}(r,\nu')=C_2(s) \frac{c r_e K'_e(r) R'(r)}{V'(r) \nu'} (\frac{\nu'_B(r)}{\nu'})^{(s+2)/2},
\end{equation}
where $r_e$ is the classical electron radius and $C_2(s)$ is a constant. It can be transformed into $\tau_{\rm SSA}(r,\nu')=\tau_{\rm SSA}(r_0,\nu')(r/r_0)^{-\alpha_B(s+2)/2+\alpha_L-1}$. The frequency, where the opacity satisfies $\tau_{\rm SSA}(r,\nu')\approx 1$, is the synchrotron self-absorption turnover frequency $\nu'_{\rm m}(r)$ at distance $r$. Therefore, we have 
\begin{equation}\label{eq:rv_ssa}
    [\nu'_{\rm m}(r)/\nu'_{\rm m}(r_0)]^{-(s+4)/2}(r/r_0)^{-\alpha_B(s+2)/2+\alpha_L-1}=1 
\end{equation}
where $\nu'_{\rm m}(r_0)$ is the turnover frequency caused by absorption at $r_0$ given as
\begin{equation}\label{eq:nu_m0}
    \nu'_{\rm m}(r_0)= [C_2(s) \frac{c r_e K'_e(r_0) R'(r_0)}{V'(r_0)} \nu'_B(r_0)^{(s+2)/2}]^{\frac{2}{s+4}}.
\end{equation}

To estimate the total radiation of blazars in the low state, we need to estimate the number of blobs produced at various locations within the jet, averaged over time. Considering that the generation and disappearance of blobs reach a dynamic equilibrium, the number of quasi-stably existing blobs at a distance from $r$ to $r+\Delta r$ is evaluated as $N_{\rm blob}(r)\approx A (r/r_0)^{-\alpha_{\rm blob}} \tau_{\rm blob}(r) \Delta r$, where $A=\dot{N}/\int(r/r_0)^{-\alpha_{\rm blob}}dr$ is a normalization and $\tau_{\rm blob}(r)=2R'(r)/c\delta_{\rm D}$ is the duration of active phase of a blob in the observer's frame. Then the total synchrotron radiation can be estimated as
\begin{equation}\label{eq:L_total}
    L^{\rm syn \ \prime}_{\nu, \rm total}(\nu') = \int_{r_0}^{r_0+L}\frac{L^{\rm syn \ \prime}_{\nu, \rm blob}(r, \nu') \frac{1-e^{-\tau_{\rm SSA}}}{\tau_{\rm SSA}} N_{\rm blob}(r)}{\Delta r}dr,
\end{equation}
where $L$ is the length of the jet. In the inner regions, the jet is generally considered to be more energetic with a higher magnetic field and particle density. However, the synchrotron self-absorption of radio synchrotron photons generated here is also very significant, resulting in the suppression of the radio flux in the optically thick inner regions. In the outer optically thin regions, the jet's radio emission gradually weakens as a result of a decreased magnetic field and particle density. Therefore, the radio core is considered to be the transition region where $\tau_{\rm SSA}(r,\nu') \approx 1$. For a specific radio frequency $\nu'$, the region at the distance $r_{\rm core}$ where $\tau_{\rm SSA}(r_{\rm core},\nu')\approx 1$ exhibits the highest flux intensity. From Equation~\ref{eq:rv_ssa}, we have
\begin{equation}\label{eq:r_c}
    r_{\rm core}(\nu')=r_0[\nu'/\nu'_{\rm m}(r_0)]^{-\frac{s+4}{\alpha_B(s+2)-2\alpha_L+2}}.
\end{equation}
At frequency $\nu'$, the regions with distances smaller than $r_{\rm core}(\nu')$ are optically thick and the emission from these places is negligible, so that the total emission can be evaluated as
\begin{equation}\label{eq:L_totala}
    L^{\rm syn \ \prime}_{\nu, \rm total}(\nu') \approx \int_{{\rm max}(r_0, r_{\rm core}(\nu'))}^{r_0+L}\frac{L^{\rm syn \ \prime}_{\nu, \rm blob}(r, \nu') N_{\rm blob}(r)}{\Delta r}dr.
\end{equation}
As a result of $u'_B(r)\nu'_B(r)^{(s-3)/2}\propto B'(r)^{(s+1)/2} \propto r^{-\alpha_B(s+1)/2}$ and $K'_e(r)\propto L^{\prime}_e(r)t'_{\rm inj}(r)\propto r^{\alpha_L+1}$, we have $L^{\rm syn \ \prime}_{\nu, \rm blob}(r, \nu')\propto r^{-\alpha_B(s+1)/2+\alpha_L+1}\nu'^{(1-s)/2}$. Based on Equation~\ref{eq:L_totala}, we obtain $L^{\rm syn \ \prime}_{\nu, \rm total}(\nu')\propto \int_{{\rm max}(r_0, r_{\rm core}(\nu'))}^{r_0+L} r^{\alpha}\nu'^{(1-s)/2}dr$ with $\alpha=-\alpha_{\rm blob}-\alpha_B\frac{s+1}{2}+\alpha_L+2$, and further acquire $L^{\rm syn \ \prime}_{\nu, \rm total}(\nu')\propto r_{\rm core}(\nu')^{\alpha+1}\nu'^{(1-s)/2}$ for $\nu'<\nu'_{\rm m}(r_0)$. Substituting Equation~\ref{eq:r_c}, the radio spectrum can be described as $L^{\rm syn \ \prime}_{\nu, \rm total}(\nu')\propto \nu'^{\alpha_r}$ and the radio spectral index $\alpha_r$ is
\begin{equation}\label{eq:ar}
    \alpha_r = -\frac{s-1}{2}+\frac{(s+4)[2\alpha_{\rm blob}+\alpha_B(s+1)-2\alpha_L-6]}{\alpha_B(2s+4)-4\alpha_L+4}.
\end{equation}

\subsection{Core-shift relation}

The core-shift phenomenon was first systematically proposed by \citet{1979ApJ...232...34B}, predicting that the distances of the ``cores'' of the jet are inversely proportional to the observed frequencies $r_{\rm core}\propto \nu^{-1}$. Later, \citet{1981ApJ...243..700K} generalized this to $r_{\rm core}\propto \nu^{-1/k_r}$, where the index $k_r$ depends on the configuration of the magnetic field and the electron density in the jet. This reveals that spatial offsets exist between cores observed in different energy bands, resulting from varying opacity levels in different physical environments. In this respect, the core-shift relation provides physical insight into the jet.

To study this phenomenon, \citet{1998A&A...330...79L} introduces a measure of the offset of core positions at two frequencies $\nu_1$ and $\nu_2$
\begin{equation}
    {\rm \Omega}_{r\nu}=4.85\times 10^{-9} \frac{\Delta \theta_{\rm mas} D_{\rm L}}{(1+z)^2} \frac{\nu_1^{1/k_r}\nu_2^{1/k_r}}{\nu_2^{1/k_r}-\nu_1^{1/k_r}},
\end{equation}
where $\Delta \theta_{\rm mas}$ is  the measured offset of the core positions, $D_{\rm L}$ is the luminosity distance to the source. The distance from the core to the SMBH is given as \citep{1998A&A...330...79L}
\begin{equation}
    r_{\rm core}(\nu)={\rm \Omega}_{r\nu}(\nu^{1/k_r}\sin{\theta_{\rm obs}})^{-1},
\end{equation}
where $\theta_{\rm obs}$ is the viewing angle of the jet with respect to the line of sight. Compared with equation (\ref{eq:r_c}), we are able to derive the relation between observed measurements and parameters
\begin{equation}\label{eq:kr}
    k_r = \frac{\alpha_B(s+2)-2\alpha_L+2}{s+4},
\end{equation}
\begin{equation}\label{eq:Orv}
    {\rm \Omega}_{r\nu} = r_0 \sin{\theta_{\rm obs}} \nu_{\rm m}(r_0)^{\frac{s+4}{\alpha_B(s+2)-2\alpha_L+2}}.
\end{equation}

\subsection{Power spectral density}\label{sec:PSD}

Light curves are fundamental temporal information recording activities that the jet has undergone. To study the temporal properties of blazar emission, we apply PSD analysis with fractional rms squared normalization as \citep[e.g.][]{2002MNRAS.332..231U}
\begin{equation}\label{eq:PSD}
    P(f)=\frac{2T}{\mu^2 N_{\rm data}^2} {\left| \mathscr{F}(f) \right|}^2,
\end{equation}
where $T$ is the total duration of the observation, $\mu$ is the mean flux, $N_{\rm data}$ is the number of data points, and $\mathscr{F}(f)$ is the Fourier transform of the observed light curve. With this normalization, we can estimate fractional variability contributed by a selected temporal-frequency range and compare PSDs derived from different instruments at different energy bands.

Within the multi-blob model framework, for a chosen set of physical parameters, we numerically simulate the light curves in different wavebands and obtain the corresponding PSDs based on Equation~\ref{eq:PSD}. In order to investigate the relationship between the power spectrum and the model parameters, we perform an analytic estimate of the power spectrum based on an arbitrary form of the light curve. For a blob at distance $r$, assuming that blobs at the same distance have the same light-curve shape and differ only in their start times, the light curve at $\nu$ can be given by 
\begin{equation}
    L_{\rm blob}(r,\nu,t) = L_{\nu, \rm blob}(r,\nu) l(r,\nu, t_m, t-t_m),
\end{equation}
where $L_{\nu, \rm blob}(r,\nu)$ is the peak flux that can be estimated using Equation~\ref{eq:L_blob_syn}, $l(r,\nu, t_m, t-t_m)$ is the dimensionless light curve and $t_m$ is the time at which the dissipation begins. The Fourier transform of the light curve is evaluated as 
\begin{equation}
    \mathscr{F}_{\rm blob}(r,\nu,f) = L_{\nu, \rm blob}(r,\nu) e^{-i2\pi f t_m} g(r,\nu,f),
\end{equation}
where $g(r,\nu,f)$ is the Fourier transform of $l(r,\nu, 0,t)$. Because the dissipation events are triggered randomly, we treat different blobs as statistically independent and neglect cross-correlations between them. Taking into account the impact of all blobs within the segment at distance $r$ over a fixed duration $T$, we have
\begin{equation}
    \mathscr{F}_{\rm seg}(r,\nu,f) = L_{\nu, \rm blob}(r,\nu) g(r,\nu,f) \sum^{N^T_{\rm blob}(r)}_{j=1} e^{-i2\pi f t_{m,j}},
\end{equation}\label{eq:ft_seg}
where $N^T_{\rm blob}(r) \approx A (r/r_0)^{-\alpha_{\rm blob}} T \Delta r$ is the number of blobs generated over duration $T$, $t_{m,j}$ is the start time of the $j$th blob. Then we have $|\mathscr{F}_{\rm seg}(r,\nu,f)|^2=L^2_{\nu, \rm blob}(r,\nu) |g(r,\nu,f)|^2[N^T_{\rm blob}(r)+\sum_{j\neq k} e^{-i2\pi f (t_{m,j}-t_{m,k})}]$. Since the starting times of each dissipation event are randomly distributed, the expectation is 
\begin{equation}
    \mathbb{E}_{t_{m}}[|\mathscr{F}_{\rm seg}|^2](r,\nu,f)=L^2_{\nu, \rm blob}(r,\nu) |g(r,\nu,f)|^2[N^T_{\rm blob}(r)+N^T_{\rm blob}(r)(N^T_{\rm blob}(r)-1){\rm sinc}^2(\pi Tf)].
\end{equation}
The off-diagonal terms contain a sinc function, which vanishes at the minimum observable frequency $f_{\rm min}=1/T$. Therefore, the expectation is dominated by diagonal terms. For a finite duration $T$, the power spectrum is $S_{\rm seg}=\left\langle|\mathscr{F}_{\rm seg}|^2\right\rangle/T$ and fractional rms PSD is $P_{\rm seg}=S_{\rm seg}/\overline{L_{\rm seg}}^2$, where $\overline{L_{\rm seg}}(t)=N^T_{\rm blob}(r)L_{\rm blob}(r,\nu,t)\int l(r,\nu, t_m,t)dt /T$ is the mean flux of the segment at distance $r$. The expectation of the fractional rms PSD can be estimated as
\begin{equation}
    \mathbb{E}[P_{\rm seg}](r,\nu,f)=\frac{T}{N^T_{\rm blob}(r)}\frac{|g(r,\nu,f)|^2}{[\int l(r,\nu, t_m,t)dt]^2}.
\end{equation}\label{eq:P_seg}
From this estimate we find that the overall amplitude of the PSD depends only on the occurrence rate of dissipation events. Under fractional rms squared normalization, the power, which represents the variability amplitude, satisfies $P_{\rm seg}\propto T/N^T_{\rm blob}(r)\propto\dot{N}^{-1}$ for a given duration $T$. Thus, more frequent events actually lead to weaker variability. This is reflected in the fact that the more blobs generated and radiating within a given period of time, the smoother the light curve becomes, thereby reducing its fluctuations. This results in a smaller standard deviation and, consequently, lower PSD levels; conversely, fewer blobs lead to larger standard deviations and higher PSD levels. This provides us with a specific method to study the dissipation frequency within the emission region. When considering the total emission, the light curves produced at different locations are no longer identical, and the resulting PSD becomes more complex. However, we can still expect that the amplitude of the overall PSD is primarily determined by the dissipation rate $P_{\rm total}\propto\dot{N}^{-1}$. 

\section{Application}\label{sec:app}

Mrk~501 is a high-synchrotron-peaked (HSP) BL Lac object with redshift $z=0.034$. It is one of the best-understood blazars, observed to emit TeV photons \citep[e.g.][]{2022ApJ...929..125A, 2024ApJS..271...25C} and exhibit profound activity and variability at keV and TeV band \citep[e.g.][]{2023ApJS..268...20K, 2020A&A...637A..86M}. As it is relatively close to us, it can be monitored by many instruments, yielding plenty of high-quality observations \citep[e.g.][]{2024A&A...685A.117M, 2021A&A...655A..93A}. In 2009, a 4.5-month multiwavelength campaign provided a comprehensive SED \citep{2011ApJ...727..129A}, which we refer to as the typical state. Another campaign was carried out by \citet{2023ApJS..266...37A} and found that it was at a low-activity state with few noticeable flares, which provides the low-state data used here. Besides, position measurements of frequency-dependent radio cores reveal the core-shift relation of Mrk~501 \citep{2010MNRAS.402..259C} and provide information about the locations of the emitting regions. These observations allow us to perform a broad model comparison for Mrk~501.

\subsection{SEDs and core-shift fitting}\label{sec:SED&CS-fit}

\begin{figure}
\centering
\includegraphics[width=1.0\columnwidth]{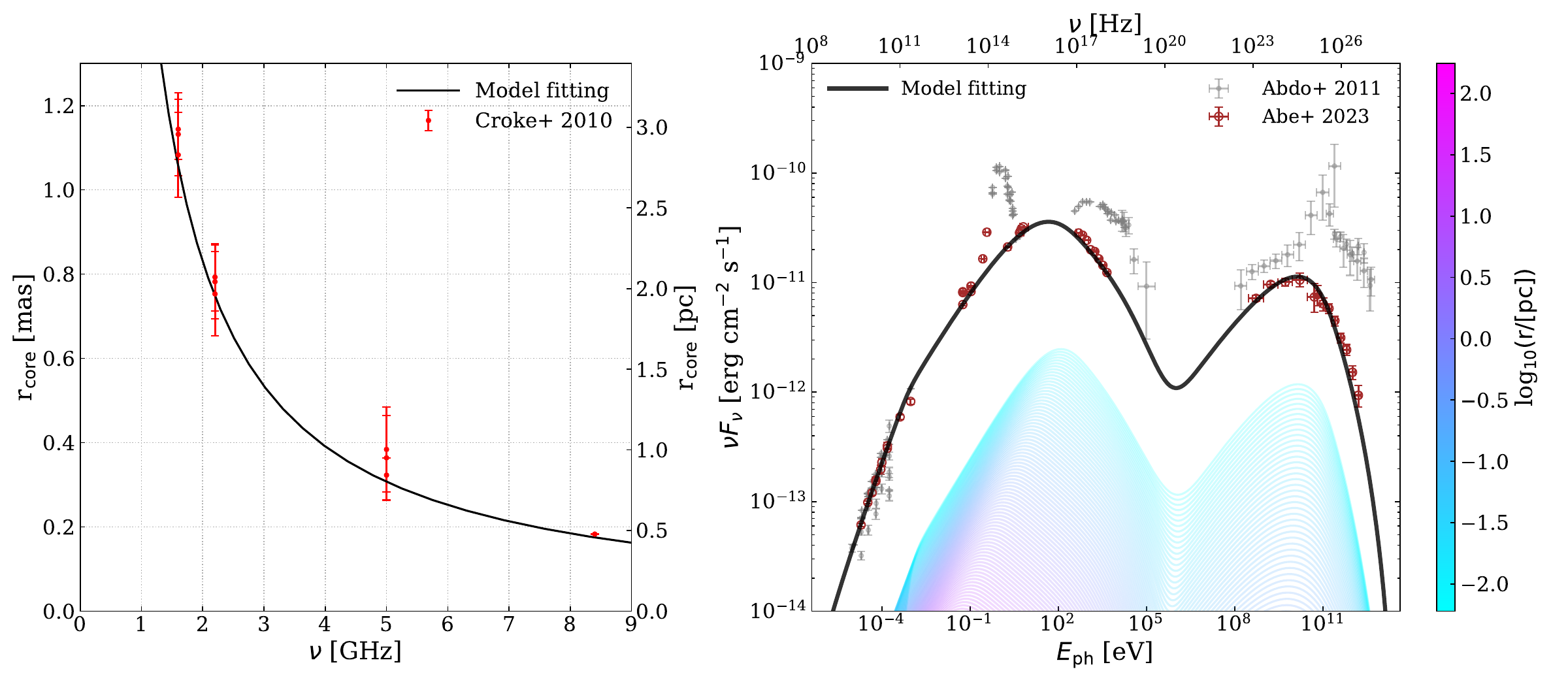}
\caption{The core-shift relation (left panel) and the SED (right panel) reproduced by the model. In the left panel, the red points show the measured positions of the frequency-dependent radio cores, and the black solid line is the model result. The corresponding core distances at 1.6, 2.2, 5.0, and 8.4\,GHz are also shown. In the right panel, the grey points show the typical state of Mrk~501 in 2009, the brown points show the low state in 2017--2019
. The black solid line is the total model emission. The colored curves show the contributions from different distances along the jet.}
\label{fig:cs&sed}
\end{figure}

\begin{table*}
\begin{minipage}[t][]{\textwidth}
\caption{Summary of model parameters for Mrk~501. The baseline model uses a single radial distribution throughout the jet. In the modified model, only some parameters in the inner jet ($r\lesssim 0.1\,{\rm pc}$) are changed, while the outer jet ($r\gtrsim 0.1\,{\rm pc}$) keeps the same properties as in the baseline model, as shown in Fig.~\ref{fig:para-mod}. The modified-model columns list the inner-jet quantities only. Parameters marked with a tilde are modified relative to the baseline model, whereas a dash indicates that the parameter is unchanged and keeps its baseline value. In the baseline model, $\kappa$ is constant, whereas in the inner jet of the modified model the radius ratio follows $\tilde{\kappa}(r)=0.31(r/r_0)^{-1/6}$.} 
\label{tab:params}
\centering
\begin{tabular}{ccccc}
\hline\hline
\multicolumn{2}{c}{Baseline Model} & \multicolumn{2}{c}{Modified Model} &	Notes	\\
\hline
Parameters & Values & Parameters & Values &  \\
\hline
$r_0$\,(pc) & 0.006 & - & - & Position of jet base \\
$B'_0$\,(G) & 0.53 & - & - & Magnetic field at $r_0$ \\
$L'_{e,0}$\,(erg\,${\rm s}^{-1}$) & $7.6\times 10^{40}$ & $\tilde{L}'_{e,0}$\,(erg\,${\rm s}^{-1}$) & $8.2\times 10^{40}$ & Injection electron luminosity at $r_0$ \\
$\kappa$ & 0.16 & $\tilde{\kappa}$ & 0.31 & Ratio of blob radius and transverse radius of jet at $r_0$ \\
$\dot{N}$\,(${\rm s}^{-1}$) & 	0.025 & $\tilde{\dot{N}}$\,(${\rm s}^{-1}$) & $1.4\times 10^{-3}$ & Total dissipation rate \\
$\delta_{\rm D}$ & 15 & - & - & Doppler factor \\
$\gamma'_{\rm min}$ & 1.0 & - & - & Minimum electron Lorentz factor \\
$\gamma'_{\rm max,1}$ & $4.0 \times 10^4$ & - & - & Lower limit of maximum electron Lorentz factor \\
$\gamma'_{\rm max,2}$ & $5.0 \times 10^6$ & - & - & Upper limit of maximum electron Lorentz factor \\
$s$ & 2.1 & - & - & Spectral index of electron energy distribution \\
$\alpha_{\rm blob}$ & 	2.0 & $\tilde{\alpha}_{\rm blob}$ & 1.5 & Index of dissipation probability $p(r)$ \\
$\alpha_B$ & 1.0 & - & - & Index of magnetic field distribution $B'(r)$ \\
$\alpha_L$ & 0.3 & $\tilde{\alpha}_L$ & 0.13 & Index of injection luminosity distribution $L'_{\rm e}(r)$ \\
$\alpha_{\gamma}$ & 2.3 & - & - & Index of blobs distribution with $\gamma'_{\rm max}$ \\
\hline	
\hline					
\end{tabular}
\end{minipage}
\end{table*}

For Mrk~501, we adopt a viewing angle $\theta_{\rm obs}\approx 15^{\circ}$ between the jet axis and the line of sight \citep{2004ApJ...600..127G}, and a jet half-opening angle of $5^{\circ}$. The measured core-position offset is $\Delta \theta_{\rm mas}=1.9\,{\rm mas}\,(\nu^{-1/0.91}-8.41^{-1/0.91})$, which gives $k_{r,{\rm obs}}=0.91\pm0.22$ \citep{2010MNRAS.402..259C}. Using Equations~(\ref{eq:ar}) and (\ref{eq:kr}), we obtain $\alpha_{\rm blob}\approx 2.0$ and $\alpha_L\approx0.3$. This estimate is mainly constrained by the 1.6--8.4\,GHz core-shift data, which correspond to distances of about 0.5--3.0\,pc.

The core-shift fit is shown in the left panel of Fig.~\ref{fig:cs&sed}. The red points are the measured core positions, and the black solid line is the model result. The multiwavelength SED fit is shown in the right panel. The model parameters are listed in Table~\ref{tab:params}. The 2017--2019 low state (brown points) is much fainter than the 2009 typical state (grey points), especially in the X-ray and $\gamma$-ray bands. Overall, the model reproduces the core-shift measurements, the flat radio spectrum, and the low-state SED. According to these parameters, the local generation rate is $\dot{N}_r=p(r)\Delta r\propto r^{-1.0}$. Thus, the model predicts that more blobs are generated closer to the black hole. If the blob lifetime $t_{\rm blob}$ scales with its size, the number of blobs coexisting in a given radial bin is $N_{\rm blob}\approx p(r)t_{\rm blob}\Delta r\propto r^0$. The electron density profile is $n'_e(r) \propto L'_e(r)t'_{\rm inj}(r)/V'_{\rm blob}\propto r^{-1.7}$.

\subsection{PSD simulation}\label{sec:psd-sim}

\begin{figure}
\centering
\includegraphics[width=1.0\columnwidth]{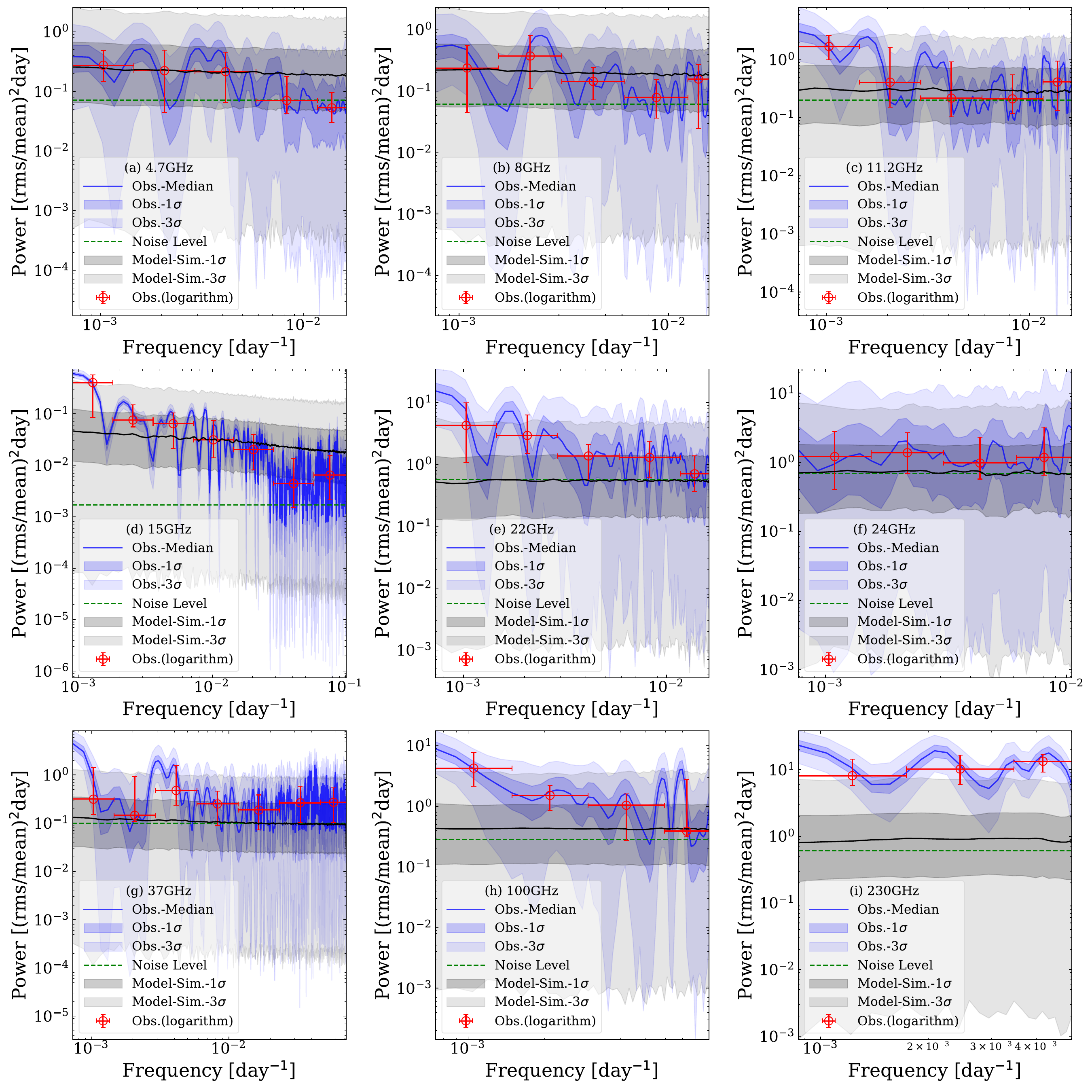}
\caption{Multiwavelength PSDs predicted by the baseline model. The panels show PSDs derived from long-term multiwavelength observations. The blue solid lines show the median PSDs obtained from Monte Carlo simulations of the observed light curves, and the blue shaded regions show the 1$\sigma$ and 3$\sigma$ ranges. The red points show the PSDs rebinned in logarithmic frequency bins. The black solid lines and shaded regions show the model PSDs after applying the same sampling pattern, window function, and observational uncertainties as in the data. The green dashed lines show the white-noise level from observational uncertainties. The bands are: (a) RATAN 4.7\,GHz, (b) Medicina 8\,GHz, (c) RATAN 11.2\,GHz, (d) OVRO 15\,GHz, (e) RATAN 22\,GHz, (f) Medicina 24\,GHz, (g) Mets\"ahovi 37\,GHz, (h) IRAM 100\,GHz, and (i) IRAM 230\,GHz.}
\label{fig:PSD}
\end{figure}

\begin{figure}
\ContinuedFloat
\centering
\includegraphics[width=1.0\columnwidth]{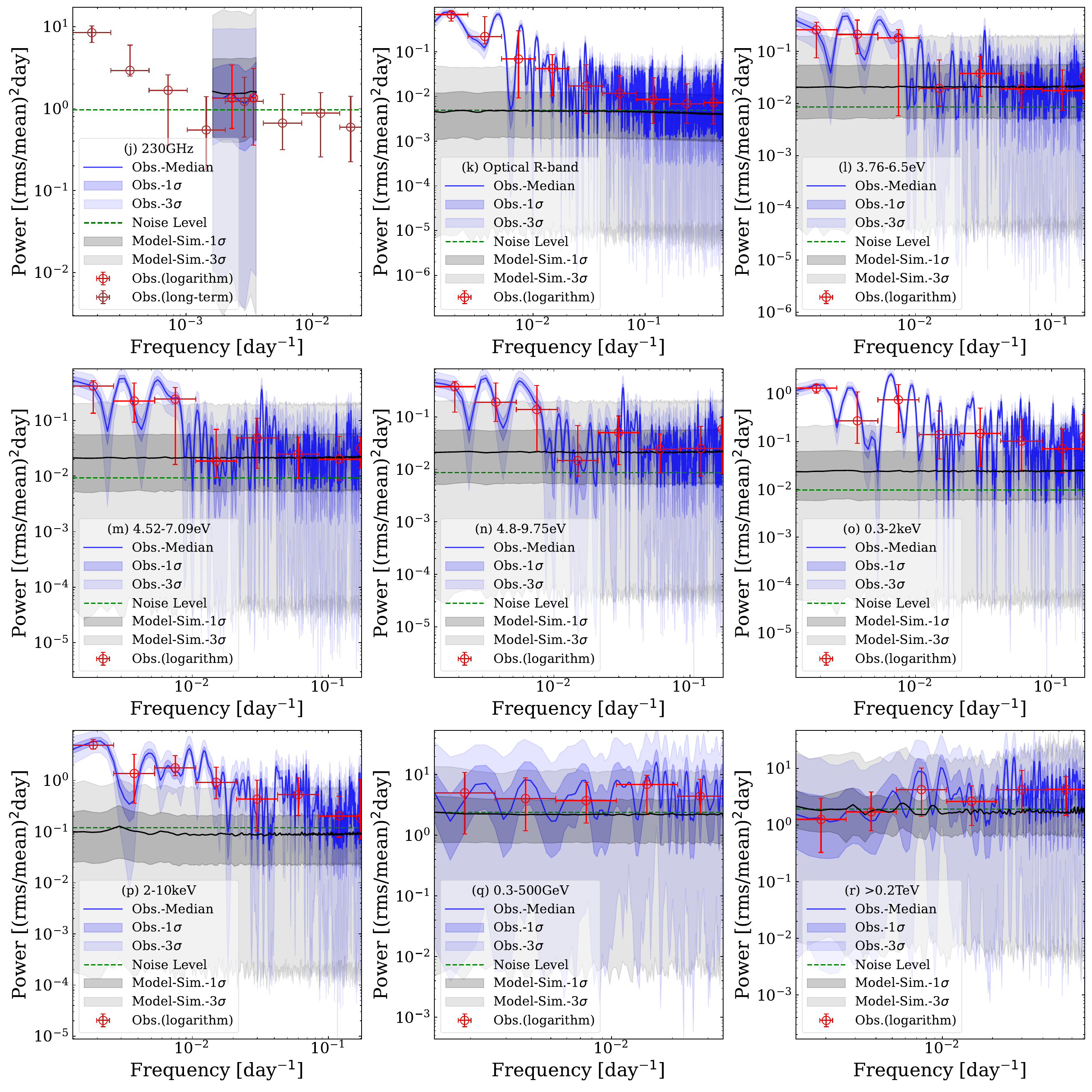}
\caption{Multiwavelength PSDs predicted by the baseline model. The remaining bands are: (j) SMA 230\,GHz, (k) GASP-WEBT and Tuorla optical $R$ band, (l) Swift-UVOT W1 (3.76--6.5~eV), (m) Swift-UVOT M2 (4.52--7.09~eV), (n) Swift-UVOT W2 (4.8--9.75~eV), (o) Swift-XRT (0.3--2~keV), (p) Swift-XRT (2--10~keV), (q) Fermi-LAT (0.3--500~GeV), and (r) MAGIC ($>$0.2~TeV).}{(continued)}
\end{figure}

Using the low-state data of \citet{2023ApJS..266...37A}, we first compute PSDs for the observed light curves and compare them with the model. The best SED and core-shift fit gives a total dissipation rate of $\dot{N}=0.025~{\rm s^{-1}}$. The PSD comparison is shown in Fig.~\ref{fig:PSD}. We compute the PSDs with the Lomb--Scargle periodogram \citep{1976Ap&SS..39..447L, 1982ApJ...263..835S}. We then use Monte Carlo simulations to include the effects of irregular sampling and observational uncertainties. The blue curves and shaded regions show the median, 1$\sigma$, and 3$\sigma$ ranges of the observed PSD estimates. The red points show the same PSDs after rebinning in logarithmic frequency bins. For the model, we first generate light curves on a uniform time grid and then resample them with the same window function as the data. We use 10,000 realizations to reduce the random effects of blob generation and sampling. Finally, we add observational uncertainties to the simulated light curves, which gives the black curves and shaded regions in Fig.~\ref{fig:PSD}.

The baseline model gives a reasonable PSD amplitude up to about 15\,GHz. At higher radio frequencies, the intrinsic model PSD is lower than the observed PSD, and the discrepancy increases with frequency. A similar discrepancy also appears in the optical, X-ray, and $\gamma$-ray bands. The reason why the discrepancy becomes larger at higher frequencies can be understood from the baseline model itself. Higher-frequency radio emission comes from regions closer to the jet base. In the single-distribution model, the local blob generation rate increases inward, $\dot{N}_{r}\propto r^{-1.0}$. The analytic estimate in Equation~(\ref{eq:P_seg}) shows that, under fractional-rms normalization, the PSD amplitude scales approximately as $P\propto 1/N_{\rm blob}^{T}$, or equivalently $P\propto \dot{N}^{-1}$, for a population of statistically independent radiating blobs. Therefore, a region containing more independent blobs produces a smoother light curve and a smaller fractional PSD. The baseline model consequently predicts weaker PSDs at higher frequencies, because these bands are dominated by inner regions with more blobs. The observations show the opposite trend, with the variability becoming stronger at higher frequencies. This indicates that the single radial distribution of effective radiating-blob properties is not sufficient.

The core-shift relation mainly constrains the jet region traced by the 1.6--8.4\,GHz cores, corresponding to distances of about 0.5--3\,pc. The agreement at 4.7 and 8\,GHz shows that the baseline model works well in this outer radio region. The agreement at 11.2 and 15\,GHz is still acceptable. However, the larger discrepancies above 15\,GHz suggest that the inner jet, at distances smaller than about 0.25\,pc, does not follow the same parameter distributions as the outer jet.

The number of radiating blobs is the parameter that most directly controls the PSD amplitude. The high observed PSDs therefore suggest that the number of blobs contributing to the inner-jet emission is much smaller than expected in the single-distribution model. For example, the fitted baseline model predicts a local dissipation rate of about $1.4\times10^{-4}(r/0.1\,{\rm pc})^{-1}~{\rm s^{-1}}$. At the 100 and 230\,GHz cores, located at about 0.03 and 0.01\,pc, the predicted rates are $4.7\times10^{-4}~{\rm s^{-1}}$ and $1.4\times10^{-3}~{\rm s^{-1}}$, respectively. To raise the model PSD amplitudes to the observed level, the effective rate of radiating blobs at the 0.01\,pc scale needs to be reduced to roughly $10^{-6}~{\rm s^{-1}}$. The optical to $\gamma$-ray emission is also expected to come mainly from the innermost region near $r_0$, and its observed PSD exceeds the unmodified model by about three orders of magnitude. This again points to an effective radiating-blob rate of order $10^{-6}~{\rm s^{-1}}$ in the innermost jet.

\subsection{PSD fitting}\label{sec:psd-fit}

\begin{figure}
\centering
\includegraphics[width=1.0\columnwidth]{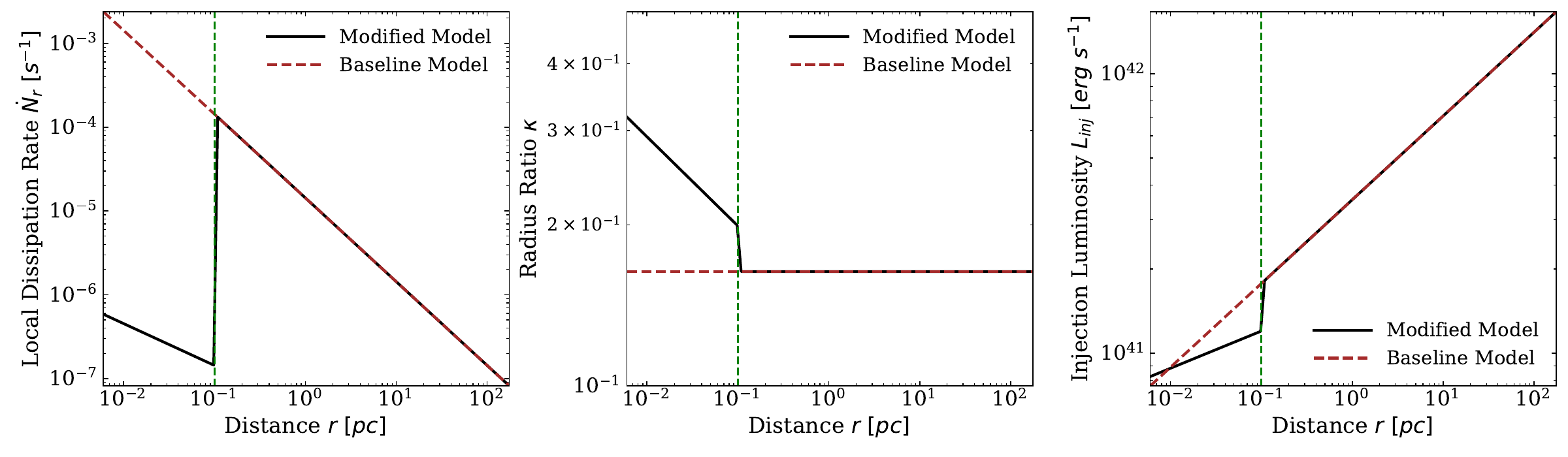}
\caption{Distance dependence of the modified model parameters, local dissipation rate $\dot{N}_r$ (left panel), the radius ratio $\kappa(r)$ (middle panel) and the injected luminosity $L'_e(r)$ (right panel). Brown dashed lines show the baseline single-distribution model used for the SED and core-shift fit. Black solid lines show the modified distributions used to fit the multiwavelength PSDs.}
\label{fig:para-mod}
\end{figure}

\begin{figure}
\centering
\includegraphics[width=1.0\columnwidth]{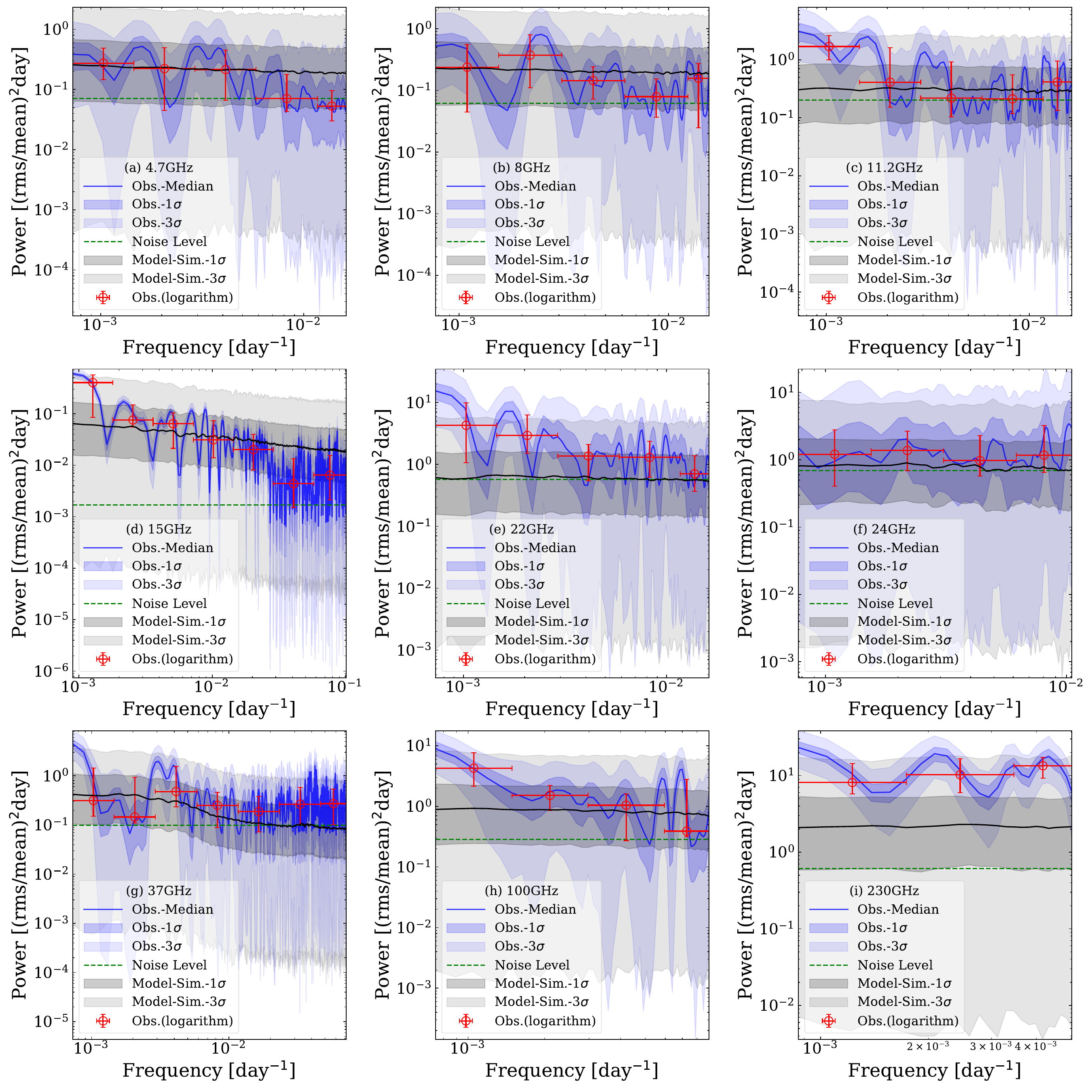}
\caption{Multiwavelength PSDs after the model modification. The legend is the same as in Fig.~\ref{fig:PSD}. Here we show the model PSDs after applying the sampling pattern, window function, and observational uncertainties of the data.}
\label{fig:PSD_2c}
\end{figure}

\begin{figure}
\ContinuedFloat
\centering
\includegraphics[width=1.0\columnwidth]{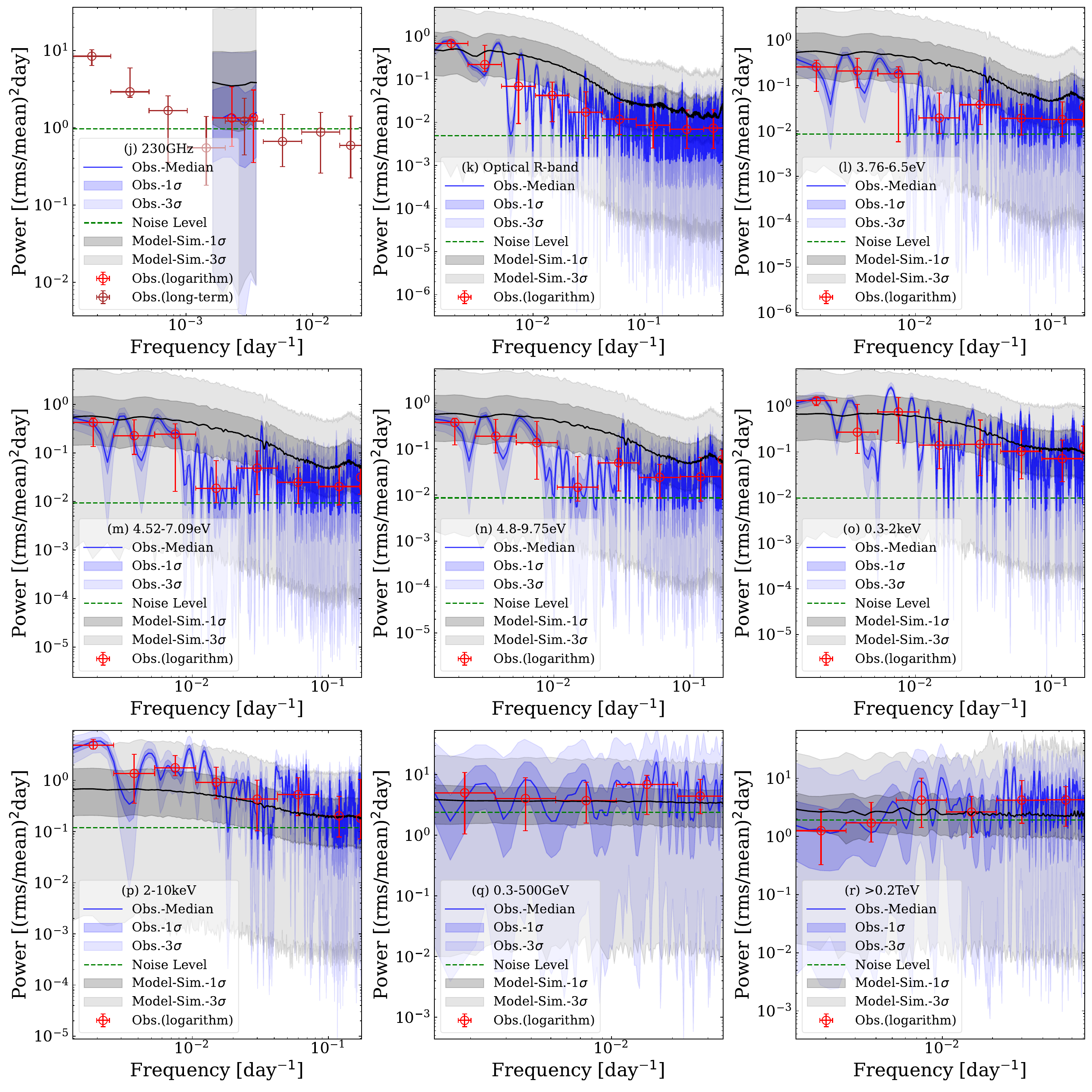}
\caption{Multiwavelength PSDs after the model modification.}{(continued)}
\end{figure}

The multiwavelength PSDs cannot be reproduced if all physical quantities follow a single radial distribution. Thus, we keep the successful SED and core-shift fit as the baseline, and then modify the model parameters in the inner jet to match the PSDs. This modified model should not be regarded as a unique solution. It is a phenomenological example showing what kind of change is required by the variability data.

The high PSD amplitudes require fewer radiating blobs. At the same time, the mean flux must remain nearly unchanged, so the remaining blobs must be larger, brighter, and/or longer-lived. We describe this behavior with a simple blob-merging picture. Note that this modified model does not attempt to follow the full nonlinear dynamics of blob interactions. Instead, it parameterizes the population of blobs that survive long enough to dominate the observed radiation. In this sense, blob merging reduces the effective number of independent radiating structures, even though the initial production of small-scale dissipative structures may still be frequent. The PSDs may reflect the effective radiating-blob population rather than the microscopic generation rate of all initially formed blobs.

To derive the inner jet parameters, we assume that the total blob volume and density are conserved during merging. It leads to $\tilde{V}'_{\rm blob}(r)/V'_{\rm blob}(r)=[\tilde{N}_{\rm blob}(r)/N_{\rm blob}(r)]^{-1}$, where a tilde denotes a modified quantity. This gives $\tilde{\kappa}(r)\propto \tilde{N}_{\rm blob}(r)^{-1/3}$. Some newly produced blobs may also merge with blobs that are already radiating. This process acts as continued particle injection and extends the lifetime of the radiating blob. We write $t'_{\rm inj}(r)=\tau_{\rm inj}(r)R'(r)/c$, with $\tau_{\rm inj}=1$ in the baseline model, and assume that the blob lifetime scales as $t_{\rm blob}\propto\tau_{\rm inj}$. In this simple merging picture, $\tilde{\tau}_{\rm inj}(r)\propto\tilde{N}_{\rm blob}(r)^{-1}$.

For a segment at distance $r$, the synchrotron luminosity can be estimated as $L^{{\rm syn}\,\prime}_{\nu,{\rm seg}}(r)\propto L'_e(r)t'_{\rm cool}(r,\gamma')t'_{\rm blob}(r)\dot{N}_r(r)$. The SSC luminosity can be estimated as $L^{{\rm SSC}\,\prime}_{\nu,{\rm seg}}(r)\propto L'_e(r)^2 t'_{\rm cool}(r,\gamma')^2 R'(r)^{-2}t'_{\rm blob}(r)\dot{N}_r(r)$. In the present fit, the magnetic field is weak enough so that adiabatic losses dominate the cooling, i.e., $t'_{\rm cool}(r,\gamma')\approx R'(r)/c$. The above scalings then become $L^{{\rm syn}\,\prime}_{\nu,{\rm seg}}(r)\propto L'_e(r)\kappa(r)^2\tau_{\rm inj}(r)\dot{N}_r(r)$ and $L^{{\rm SSC}\,\prime}_{\nu,{\rm seg}}(r)\propto L'_e(r)^2\kappa(r)\tau_{\rm inj}(r)\dot{N}_r(r)$. In the blob-merging scenario, one simple way to keep both the synchrotron and SSC fluxes approximately unchanged is
\begin{equation}\label{eq:mod}
    \left\{
    \begin{aligned}
        &\frac{\tilde{L}'_e(r)}{L'_e(r)}=\frac{\tilde{\kappa}(r)}{\kappa(r)},\\
        &\frac{\widetilde{\dot{N}}_r(r)}{\dot{N}_r(r)} = \left[\frac{\tilde{\kappa}(r)}{\kappa(r)}\right]^{-3}\left[\frac{\tilde{\tau}_{\rm inj}(r)}{\tau_{\rm inj}(r)}\right]^{-1}.
    \end{aligned}
    \right.
\end{equation}
This prescription keeps the time-averaged SED close to the baseline fit while changing the variability amplitude. It is only one possible way to maintain an approximately unchanged overall emission level. Other choices of inner-jet parameters may lead to different observational signatures. Future core-shift measurements in the sub-millimeter/millimeter bands, i.e., $\sim0.1$--1\,THz, can test this picture and further constrain the model parameters.

The adopted parameterization is shown in Fig.~\ref{fig:para-mod}. Guided by the PSD comparison, we place the transition at 0.1\,pc, inside the region corresponding to the 15\,GHz core. In the inner jet, we set $\tilde{\tau}_{\rm inj}=10^3$ so that the radiating blobs are long-lived. For $r\gtrsim0.1~{\rm pc}$, we assume that the merging effect weakens and keeps $\tau_{\rm inj}=1$. We also let $\tilde{\kappa}(r)$ and $\tilde{L}'_e(r)$ increase toward smaller radii as $r^{-1/6}$ relative to the baseline model. This increases the contribution of the inner jet to the variability. The local dissipation rate $\tilde{\dot{N}}_r$ is then adjusted according to Equation~(\ref{eq:mod}). With these changes, the total effective dissipation rate becomes $\tilde{\dot{N}}\simeq1.4\times10^{-3}~{\rm s^{-1}}$, about one order of magnitude lower than in the baseline model. This decrease mainly reflects the reduced number of radiating blobs in the inner jet. The resulting PSDs are shown in Fig.~\ref{fig:PSD_2c}. The modified model produces stronger red-noise variability and matches the observed PSDs within the 1$\sigma$ range over a broad energy range, from GHz radio bands to TeV $\gamma$-rays.

It should be noted that the 230\,GHz PSD has large uncertainty because the number of data points is small. There are 11 low-state IRAM points and 4 SMA points during the same period. The IRAM PSD is much higher than the PSDs at other radio frequencies, and even higher than those in the X-ray and $\gamma$-ray bands. This high value may partly come from statistical fluctuations. If the intrinsic PSD is stable over a long time, any observed light curve is still only one realization of the underlying stochastic process. Sparse sampling can therefore produce an upward fluctuation in the measured PSD. Nevertheless, our model is marginally consistent with the IRAM result within the 3$\sigma$ range, indicating that such high-variability realizations are possible. The SMA PSD is more moderate and is comparable to the PSDs in the other radio bands. We also compare with long-term SMA monitoring at 230\,GHz. These data are not strictly simultaneous with the low state, but they provide an additional check on the intrinsic PSD. As shown by panels (i) and (j) of Fig.~\ref{fig:PSD_2c}, both the simultaneous and long-term SMA PSDs are about one order of magnitude lower than the IRAM PSD over the same temporal-frequency range, and the model reproduces the SMA results within the 1$\sigma$ range.

\section{Discussion}

\subsection{Prospects for the inner jet}

\begin{figure}
\centering
\includegraphics[width=0.45\columnwidth]{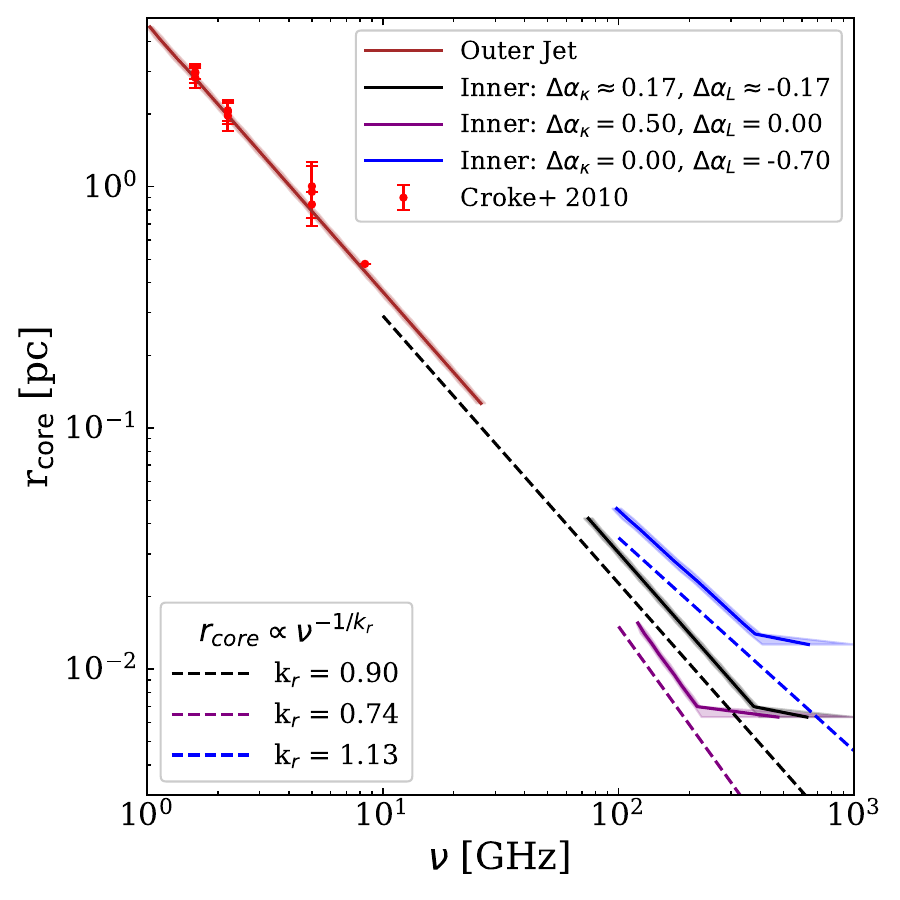}
\caption{Core-shift relations in the inner jet. The black, blue, purple solid lines represent the core-shift relations in the inner jet with different parameter distributions. The black solid line is the case mentioned in Section~\ref{sec:psd-fit}. For visual clarity, the $r_{\rm core}$ values of the blue line have been multiplied by a factor of 2. The brown solid line is the core-shift relation in the outer jet (and also in the baseline model) for reference. The dashed lines show the analytical expectation of the core-shift index, which are in good agreement with the numerical results (solid lines). All regions become optically thin above several hundred GHz. Consequently, the emission is dominated by the innermost region, causing flattening of the lines at high frequencies.}
\label{fig:inner_cs}
\end{figure}

The model discussed in Section~\ref{sec:psd-fit} represents one possible scenario that can reproduce the overall SED. At present, the observed variation in the amplitudes of the multiwavelength PSDs provides constraints only on the distribution of the number of radiating blobs, $\alpha^{\rm in}_{\rm blob}=1.5$. When the ratio between the blob radius and the jet radius is also distance-dependent, $\kappa(r)\propto r^{-\alpha_\kappa}$, the corresponding distributions of this ratio and the injection luminosity can differ from those in the outer jet region in a variety of ways. Different parameter distributions lead to different radiation profiles in the inner jet. Such modifications are expected to manifest themselves in the core-shift relation of the inner jet, as indicated by Equation~\ref{eq:mod_ar&kr}. Figure~\ref{fig:inner_cs} presents the core-shift relations in the inner jet predicted by different parameter distributions. $\Delta \alpha = \alpha^{\rm in} - \alpha^{\rm out}$ denotes the deviation of the modified parameters of distribution in the inner jet relative to those in the outer jet. The numerical calculations (solid lines) are in good agreement with the theoretical predictions (dashed lines). We did not show the expected core-shift behavior around the transition radius ($r\sim 0.1$\,pc), because the expected behavior in this gap highly depends on how parameters transits from the outer jet and the inner jet radius. In the present work, we simply assume an abrupt change of the parameters. As a result, the emission over a relatively broad frequency range is dominated by the region around 0.1\,pc. The black curve corresponds to the case adopted in Section~\ref{sec:psd-fit}, in which the radiation of every segment is assumed to remain unchanged. Consequently, the core-shift index remains identical to that of the outer jet. When $\Delta \alpha_L \neq  -\Delta \alpha_\kappa$, the core-shift relation in the inner jet exhibits significant deviations, as illustrated by the blue and purple lines. In addition, the integrated radio spectrum is also expected to be affected slightly, mainly at high radio frequencies ($\gtrsim $ 10\,GHz). Therefore, future core-shift measurements at higher radio frequencies, together with high-precision radio spectral observations, are expected to provide stronger constraints on the physical properties of the inner jet and offer further insight into its underlying structure and dynamics.

\subsection{Dissipation mechanism}

Magnetic reconnection is a promising energy-dissipation mechanism in blazar jets because it can accelerate particles efficiently \citep[e.g.][]{2014ApJ...783L..21S, 2018MNRAS.481.5687P}. Near the SMBH, jets may be launched through the Blandford--Znajek mechanism \citep{1977MNRAS.179..433B, Yang2024}. The outflow is then expected to be Poynting-flux dominated, and the strong magnetic field can be unstable. Current-driven kink instabilities \citep[e.g.][]{2008A&A...492..621M, 2017MNRAS.469.4957B} or Kelvin--Helmholtz instabilities in velocity shear layers \citep[e.g.][]{2021ApJ...907L..44S, 2025ApJ...979..199S} may trigger reconnection. In a reconnection layer, tearing instability can form many magnetically confined plasmoids \citep[e.g.][]{2010PhRvL.105w5002U}. These plasmoids are natural candidates for the emitting blobs in our model.

In this picture, particles accelerated by magnetic reconnection are confined in plasmoids and radiate there. The blob-merging effect discussed in Section~\ref{sec:psd-fit} can be interpreted as the coalescence of magnetic islands in the reconnection layer \citep[e.g.][]{2007PhPl...14g2303D, 2018MNRAS.481.5687P}. Both the reconnection rate and the merger rate are related to the local Alfv\'en speed \citep[e.g.][]{2016MNRAS.462...48S}. As the jet propagates outward, electromagnetic energy is gradually converted into bulk kinetic energy, the bulk Lorentz factor increases, and the magnetization decreases \citep[e.g.][]{2019MNRAS.490.2200C}. In the inner jet, the higher magnetization and faster Alfv\'en speed may make plasmoid production and coalescence more efficient \citep[e.g.][]{2007PhPl...14j0703L, 2009PhRvL.103j5004S}. Many small plasmoids can then merge into larger and more energetic structures, sometimes called ``monster plasmoids'' \citep{2013ApJ...775...50T}. If coalescence is efficient, newly formed plasmoids may merge before reaching a steady radiative state. The number of radiating plasmoids can then be much smaller than the number of initially generated plasmoids, while the surviving plasmoids are larger and longer-lived. This provides a possible explanation for the stronger variability required by the high frequency PSDs. In the outer jet, the lower magnetization reduces the coalescence rate. Primary plasmoids can radiate before strong merging occurs, so the number of radiating plasmoids is closer to the generation rate assumed in the baseline model.

The energy dependence of the PSDs may also indicate a change in the dominant dissipation mechanism with distance. In a magnetically driven jet, magnetic reconnection is expected to be important at small radii, where the jet is Poynting-flux dominated. On larger scales, part of the magnetic energy is converted into the bulk kinetic energy of matter. Shocks may then become more important, such as internal shocks caused by collisions between inhomogeneities in the jet \citep[e.g.][]{2010ApJ...711..445B} or recollimation shocks produced when the expanding jet is confined by the surrounding pressure field. Other acceleration processes may also contribute \citep{2002ChJAA...2....1W, 2006A&A...453...47K, 2011ApJ...739...66T}. The jet is likely governed by several processes at the same time, and the dominant process may change with distance. Future multiwavelength monitoring with denser sampling will be useful for testing this picture.

\section{Conclusion}\label{sec:conclusion}
The flat radio spectrum and the core-shift effect are two basic signatures of blazar jets. In this work, we model the multiwavelength SEDs of Mrk~501, including its flat radio spectrum, together with the observed radio core-shift relation and multi-frequency variability information. With this spectro-timing-astrometric jet modeling, we can constrain not only the radial distributions of magnetic fields and relativistic particles traced by the radio spectrum and core-shift relation, but also the effective number, size, and lifetime of radiating blobs traced by multiwavelength variability. This combines spectral, spatial, and temporal constraints in the multi-zone stochastic dissipation framework.

For Mrk~501 in the 2017--2019 low state, the baseline model fits the SED and core-shift relation well. It gives a blob generation rate $\dot{N}_r\propto r^{-1.0}$ and an electron density profile $n'_e(r)\propto r^{-1.7}$ under the assumption of conserved magnetic power along the jet. However, when the same single radial distribution is used to simulate the multiwavelength PSDs, the model underpredicts the variability at high radio frequencies ($>15$\,GHz) and at optical to $\gamma$-ray energies. This is because the baseline model produces many independent blobs in the inner jet. Many independent events smooth the light curve and reduce the fractional PSD amplitude. A smaller number of effective radiating blobs in the inner jet is required to reproduce higher-frequency PSD data.

We then introduce a simple blob-merging scenario and allow the inner and outer jet regions to have different effective parameter distributions. In the inner jet ($\lesssim 0.1~{\rm pc}$), radiation comes from fewer, larger, more energetic, and longer-lived blobs. In the outer jet ($\gtrsim 0.1~{\rm pc}$), the blob properties approach the baseline distribution constrained by the radio spectrum and core-shift data. With these modifications, the model reproduces the multiwavelength PSD amplitudes from GHz radio bands to TeV $\gamma$-rays within the observational uncertainties while keeping the SED approximately unchanged. This phenomenological solution is not unique, but it demonstrates that the variability data require an effective inner-jet radiating population different from that inferred by extrapolating the outer-jet distribution inward.

The detailed nature of jet dissipation remains uncertain and should be tested by future observations and numerical simulations. A possible physical origin is magnetic reconnection with efficient plasmoid coalescence in the highly magnetized inner jet. As the jet expands and the magnetization decreases, coalescence becomes less important and the radiating blob population approaches the generated blob population. Other dissipation channels, such as shocks, may also contribute at larger distances. The transition appears to occur on sub-parsec scales, around $r\sim0.1\,{\rm pc}\,(M_{\rm BH}/10^9M_\odot)\sim10^3R_g$. Dense multiwavelength monitoring and millimeter/sub-millimeter VLBI observations will be important for understanding this scale-dependent dissipation picture.

\begin{acknowledgments}

This study is funded by Basic Research Program of Jiangsu under grant No. BK20250059, and National Natural Science Foundation of China under grants No.~12393852 and No.~12333006.
\end{acknowledgments}

%





\clearpage
\appendix
\restartappendixnumbering

\section{Observational limitations and noise}

\begin{figure}
\centering
\includegraphics[width=0.90\columnwidth]{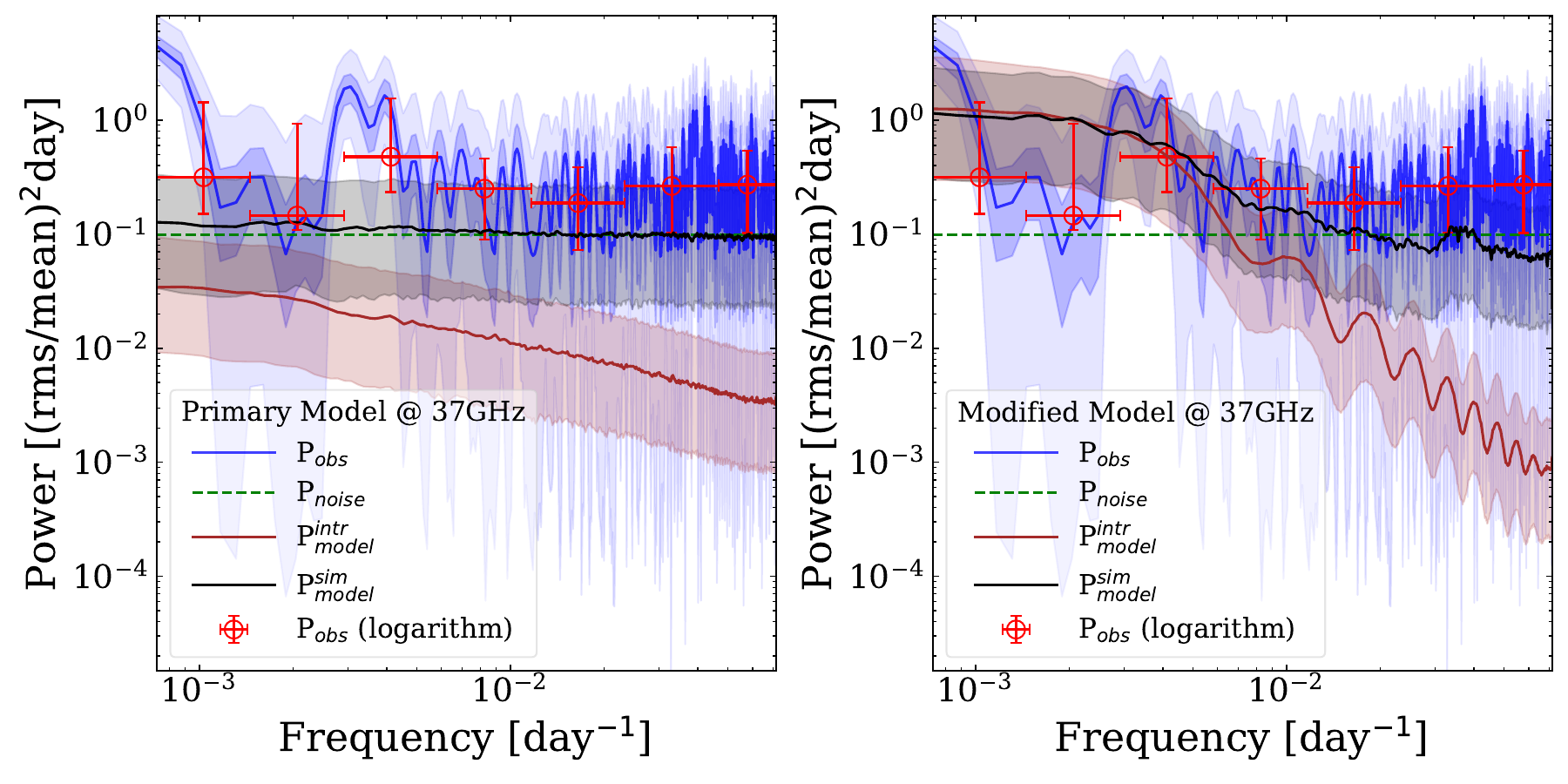}
\caption{Effects of observational limitations on PSDs. Simulated PSDs at 37\,GHz before (left panel) and after (right panel) the model modification. The brown line represents the intrinsic model PSD, while the black line shows the model PSD after including the effects of observational uncertainties, the sampling pattern and window function.}
\label{fig:psd_obs-limit}
\end{figure}

For a given light curve $L(t)$ over a finite duration $T$, its power spectrum is $S_{L}(f)=\left\langle |\mathscr{F}_L(f)|^2\right\rangle/T$ and fractional rms PSD is $P_L(f)=S_{L}(f)/\mu_L^2$, where $\mu_L=\overline{L(t)}$. It can represent the fractional variability of the light curve as
\begin{equation}
    F_{\rm var} = \sqrt{\frac{\sigma^2_L}{\mu^2_L}}=\sqrt{\int P_L(f)df},
\end{equation}
where $\sigma^2_L$ is the variance of the light curve. In real observations, limited detective efficiency and fluctuations from background introduce observational uncertainties $\sigma_{\rm err}$. This adds a white-noise component to the observed PSD as
\begin{equation}
    P_{\rm noise}=\overline{\sigma_{\rm err}}^2 / \mu^2_L.  
\end{equation}
In addition, during observations of a target source, detectors cannot continuously monitor the source without interruption. Instead, observations are usually carried out intermittently, with observing periods separated by gaps in time. This means that the interval between the centers of two adjacent observing windows $\Delta t$ is longer than the window length itself $\Delta w$. As a result, high-frequency signals within the frequency range from $1/(2\Delta t)$ to $1/(2\Delta w)$ can be shifted into lower frequency range, which is known as aliasing \citep[e.g.][]{2002MNRAS.332..231U}. It can be approximately estimated as a constant power added to the intrinsic PSD, 
\begin{equation}
    P_{\rm aliase}=\frac{1}{\mu^2_L}\frac{1}{(2\Delta w)^{-1}-(2\Delta t)^{-1}}\int^{(2\Delta w)^{-1}}_{(2\Delta t)^{-1}} S_L(f)df.
\end{equation}
If the variability amplitude of the signal is too weak $P_L(f)<P_{\rm noise}+P_{\rm aliase}$, both of these effects can raise the observed PSD at high frequencies, making the high-frequency structure of the intrinsic PSD  difficult to measure. 

The most significant effect on the observed PSD comes from irregular sampling. Irregular sampling effectively multiplies the intrinsic continuous variability signal by a complex sampling function. Consequently, the Fourier transform of the observed light curve is given by the convolution of the intrinsic Fourier transform and the transform of the sampling function. This makes it difficult to directly recover the intrinsic PSD from the observations. Therefore, we can only apply the same sampling pattern and window function to the simulated model light curves and then compare them with the observations through Monte Carlo simulations. In this way, the influence of irregular sampling can be minimized as much as possible.

The Fig.~\ref{fig:psd_obs-limit} mainly illustrates the effects of observational limitations on the PSD. The brown line represents the intrinsic model PSD. The black line shows the model PSD after including the same sampling pattern and observational uncertainties, thereby accounting for the effects of irregular sampling, aliasing and noise. For the baseline model, the observational noise level is higher than the intrinsic model PSD, causing the simulated PSD to be dominated by observational noise. In contrast, the modified model exhibits a significantly enhanced PSD amplitude. In this case, the PSD is dominated by the intrinsic model component at low frequencies, while only at high frequencies does it become dominated by noise, similar to the observations.

\section{SEDs and core-shift relation}

In order to account for the multiwavelength PSDs, we find that the physical properties of the jet must evolve from the inner to the outer regions. When considering radial dependencies for the injection timescale, $\tau_{\rm inj}(r)\propto r^{-\alpha_\tau}$, the ratio between the blob size and the jet radius, $\kappa(r)\propto r^{-\alpha_\kappa}$, the radio spectral index given by Equation~\ref{eq:ar} and the core-shift index defined in Equation~\ref{eq:kr} are correspondingly modified to
\begin{equation}\label{eq:mod_ar&kr}
    \left\{
    \begin{split}
        & \alpha_r = -\frac{s-1}{2}+\frac{(s+4)[2\alpha_{\rm blob}+\alpha_B(s+1)-2\alpha_L+4\alpha_\kappa+2\alpha_\tau-6]}{\alpha_B(2s+4)-4\alpha_L-4\alpha_\kappa+4} \\
        & k_r = \frac{\alpha_B(s+2)-2\alpha_L-2\alpha_\kappa+2}{s+4}
    \end{split}
    \right. 
\end{equation}
The solution obtained in the baseline model corresponds to $\alpha^{\rm out}_{\rm blob}=2.0, \alpha^{\rm out}_L=0.3, \alpha^{\rm out}_\kappa=\alpha^{\rm out}_\tau=0$, which is maintained in the outer jet in the modified scenario. The observed variation in the PSD amplitudes across different wavebands supports a decreasing number of blobs towards the central region, as $\Delta \alpha_{\rm blob} = \alpha^{\rm in}_{\rm blob} - \alpha^{\rm out}_{\rm blob} = -0.5$. In a simplified blob-merging model, if the total radiation from each segment remains approximately unchanged, one obtains $\Delta \alpha_{\rm blob} = -3 \Delta \alpha_\kappa - \Delta \alpha_\tau$ and $\Delta \alpha_L =  -\Delta \alpha_\kappa$. Consequently, the radial distributions of the physical quantities in the inner jet region are fully determined, with $\alpha^{\rm in}_{\rm blob}=1.5$, $\alpha^{\rm in}_\kappa=1/6$ and $\alpha^{\rm in}_L=2/15$.

\begin{figure}
\centering
\includegraphics[width=0.9\columnwidth]{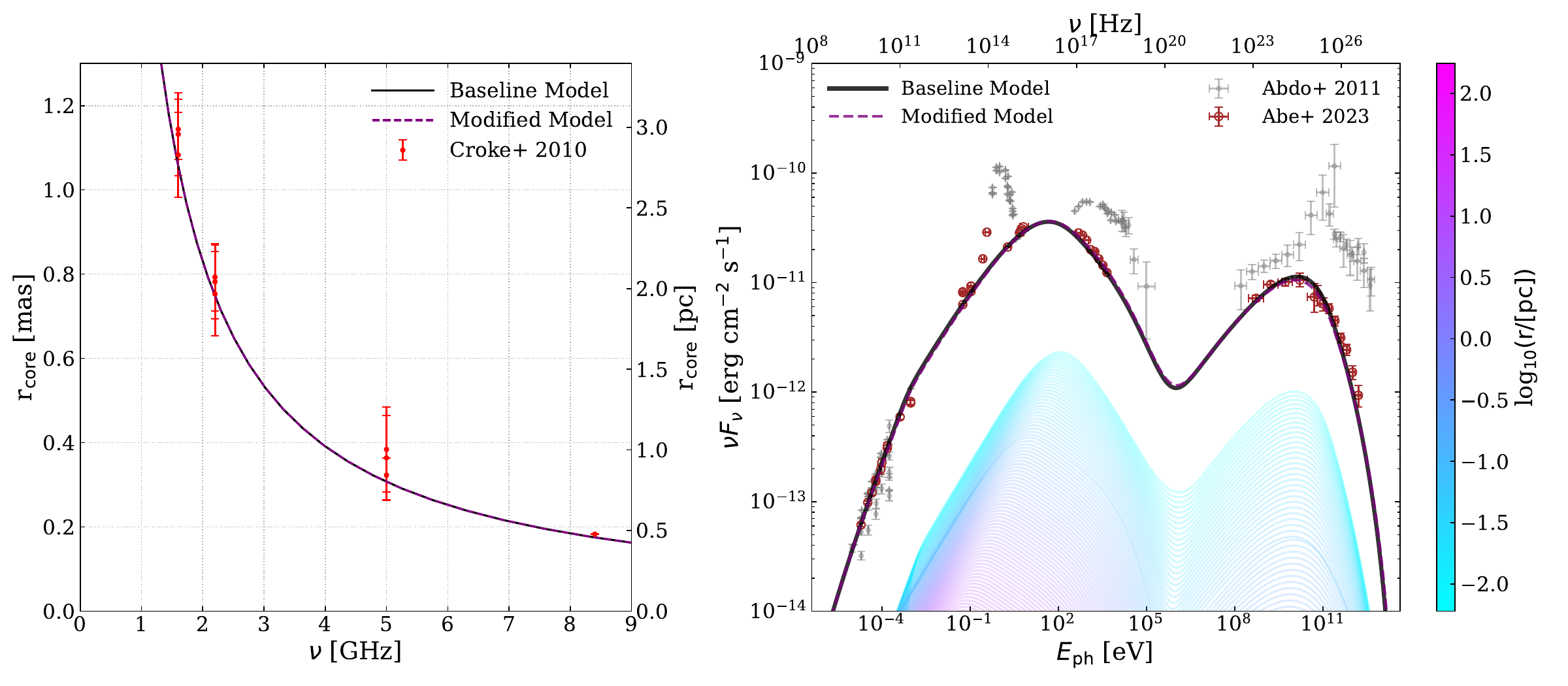}
\caption{Model SEDs and core-shift relations before (black solid lines) and after (purple dashed lines) the model modification.}
\label{fig:cs&sed_com}
\end{figure}

As shown in the left panel of Fig.~\ref{fig:cs&sed_com}, the results before and after the model adjustment are nearly identical. The dependence of the core position on radio frequency remains unchanged, with only a slight shift in the fitted curves due to the parameter modifications. For the SEDs, the parameter adjustments are performed according to the prescriptions described in Section~\ref{sec:psd-fit}, such that the synchrotron and SSC emission from each region within 0.1\,pc remain approximately unchanged. Consequently, the overall radiation level also remains nearly the same, as shown in the right panel of Fig.~\ref{fig:cs&sed_com}.

\bibliography{ref}{}
\bibliographystyle{aasjournalv7}



\end{document}